\documentclass[aps,pra,showpacs,showkeys,preprintnumbers,amsmath,amssymb]{revtex4}
\usepackage{graphicx}% Include figure files
\usepackage{dcolumn}% Align table columns on decimal point
\usepackage{bm}% bold math
\begin{document}
%%%%%%%%%%%%%%%%%%%%%%%%%%%%%%%%%%%%%%%%
\def\ibd {\it ibid.}
\def\vep{\varepsilon}
\def\bfk{{\bf k}}
\def\bfk'{{\bf k'}}
\def\bfq{{\bf q}}
\def\bfr{{\bf r}}
\def\bfp{{\bf p}}
\def\pl{\partial}
\def\tl{\tilde}
\def\zbar{{\bar z}}
\def\al{\alpha}
\def\be{\beta}
\def\gam{\gamma}
\def\lam{\lambda}
\def\bfrho{\bf \rho}
\def\nab{\bf \nabla}
\def\Delmu{\Delta \mu}
\def\beq{\begin{equation}}
\def\eeq{\end{equation}}
\def\bea{\begin{eqnarray}}
\def\eea{\end{eqnarray}}
%%%%%%%%%%%%%%%%%%%%%%%%%%%%%%%%%%%%%%%%
%\preprint{2007-MMU}
\title{Higher order terms in the condensate fraction
of a homogeneous and dilute Bose gas}
%%%%%%%%%%%%%%%%%%%%%%%%%%%%%%%%%%%%%%%%
%%%%%%%%%%%%%%%%%%%%%%%%%%%%%%%%%%%%%%%%
\author{Sang-Hoon \surname{Kim}\footnote{shkim@mmu.ac.kr}}
\affiliation{Division of Liberal Arts $\&$ Science, Mokpo National
Maritime University, Mokpo 530-729, Korea,}
%%%%%%%%%%%%%%%%%%%%%%%%%%%%%%%%%%%%%%%%
\date{\today}
\begin{abstract}
The condensate fraction of a homogeneous and dilute Bose gas
is expanded as a power series of $\sqrt{n a^3}$
as $N_0/N = 1 -c_1 (n a^3)^{1/2} -c_2 (n a^3) - c_3 (n a^3)^{3/2}\hdots.$
The coefficient $c_1$ is well-known as
$8/3 \sqrt{\pi}$, but the others are unknown yet.
Considering two-body contact interactions and applying
a canonical transformation method twice
we developed the method to obtain the higher order coefficients
analytically.
An iteration method is applied to make up a cutoff in a fluctuation term.
The coefficients ares $c_2=2(\pi - 8/3)$ and $c_3=(4/\sqrt{\pi}) \left( \pi -8/3\right)\left( 10/3-\pi \right)$.

\end{abstract}
\maketitle

%%%%%%%%%%%%%%%%%%%%%%%%%%%%%%%%%%
\section{introduction}
%%%%%%%%%%%%%%%%%%%%%%%%%%%%%%%%%%
Homogenous and dilute (HD) Bose gas has been studied for more than
a half century
since the classical paper by Bogoliubov \cite{bogol}.
Despite its simplicity and significant published research,
there still remains some unsolved fundamental problems.
The condensate fraction, or the ratio of particles of zero momentum
in the perturbed ground state,
is a very fundamental concept of a many-body Bose system.
At zero temperature, the  properties of HD gases are expressed
in terms of a single parameter, the s-wave scattering length $a$, and
 a quantum loop expansion is expressed by dilute gas parameter $n a^3 \ll 1$
where $n=N/V$ is the particle number density.

 In a HD system the condensate fraction is expressed
as \cite{huan,fett}
\begin{equation}
\frac{N_0}{N} = 1 -c_1 (n a^3)^{1/2} -c_2 (n a^3) -c_3 (n a^3)^{3/2} - \hdots.
\label{1}
\end{equation}
The coefficient $c_1$ is well-known as $c_1= 8/3 \sqrt{\pi}$, but
the others  are unknown yet.
In this paper,
 we will calculate the unknown coefficients $c_2$ and $c_3$
 of the next higher order terms analytically
using a double canonical transformation and an iteration method.

%%%%%%%%%%%%%%%%%%%%%%%%%%%%%%%%%%
\section{Bogoliubov Hamiltonian}
%%%%%%%%%%%%%%%%%%%%%%%%%%%%%%%%%%

If the interparticle interaction is
a contact potential, the effective Hamiltonian for the
identical particles of mass $m$ is written as the Bogoliubov form
\begin{equation}
\hat{H} = \sum_{ k }  \vep_k a_{k}^{\dag} a_{k} +
\frac{g}{2V} \sum_{ {{ k_{1} k_{2}}  \atop {k_{3} k_{4} }}  }
  a_{k_{1}}^{\dag}  a_{k_{2}}^{\dag} a_{k_{3}} a_{k_{4}}
  \delta_{k_{1}+k_{2},k_{3}+k_{4}},
\label{3}
\end{equation}
where $\vep_k = \hbar^2 k^2/2m$ and $g$ is the coupling constant given by $g=4
\pi \hbar^2 a/m$.
 $a_{k_i}$ and $a_{k_i}^\dagger$
satisfy the bosonic commutator relation:
$[a_k,a_{k'}^\dagger]=\delta_{k,k'}$.
Note that this Hamiltonian is valid only for the purpose of applying
first-order perturbation theory.

Since $N_0$ is a very large
number, the operators $a_0$ and $a_0^\dagger$ can be regarded as
$\sqrt{N_0}$.
 In a two-body interaction, the interacting part of the Hamiltonian
 in Eq. (\ref{3}) is
expanded to keep  the terms up to the
 order of $N_{0}^{2}, N_{0}$, and $\sqrt{N_0}$ as \cite{abri,maha,kim}
\begin{eqnarray}
{\hat{H}}_{int} &=& \frac{g}{2V} a_{0}^{\dag} a_{0}^{\dag} a_{0}
a_{0}
\nonumber \\
&  + &\frac{g}{2V} \sum_{ k \neq 0 }
 \left[ 2( a_{k}^{\dag} a_{0}^{\dag} a_{k} a_{0}
 + a_{-k}^{\dag} a_{0}^{\dag} a_{-k} a_{0} ) + a_{k}^{\dag}
a_{-k}^{\dag} a_{0} a_{0} + a_{0}^{\dag} a_{0}^{\dag} a_{k} a_{-k}
\right]
\nonumber \\
& + &\frac{g}{V} \sum_{ k , q \neq  0 }
 \left[   a_{k+q}^{\dag} a_{0}^{\dag}  a_{k} a_{q}
+ a_{k+q}^{\dag} a_{-q}^{\dag} a_{k} a_{0}  \right].
 \label{4}
\end{eqnarray}
The three-body interaction is not considered.
The last term, which is generally neglected in textbooks,
originates from the interactions of particles out of and into the
condensate and the key of the calculation.

Let's introduce a new  variable as
\begin{equation}
\gamma_{k} = \sum_{q \neq 0}  a_{k+q}^{\dag} a_{q}.
\label{5}
\end{equation}
Since the $N_0$ is expressed by the number relation
\begin{equation}
 N_0  = \hat{N} - \frac{1}{2}
\sum_{k \neq   0} ( a_{k}^{\dag} a_{k} +  a_{-k}^{\dag} a_{-k}),
\label{6}
\end{equation}
 we can rewrite the model Hamiltonian as
 a function of the density $n$ as
\begin{eqnarray}
\hat{H} &=&  \frac{1}{2} n^2 V g
 \nonumber \\
  & + &\frac{1}{2} \sum_{k \neq  0}
\left[ (\vep_k + n g) (a_{k}^{\dag} a_{k} + a_{-k}^{\dag}
a_{-k})
 + n g (a_{k}^{\dag} a_{-k}^{\dag} + a_{k} a_{-k}) \right]
  \nonumber \\
& + &\frac{ng}{\sqrt{N}}
 \sum_{k \neq  0} \gamma_k ( a_k  + a_{-k}^{\dag}  ).
\label{7}
\end{eqnarray}

%%%%%%%%%%%%%%%%%%%%%%%%%%%%%%%%%%
\section{Double canonical transformation}
%%%%%%%%%%%%%%%%%%%%%%%%%%%%%%%%%%

Next, we apply the conventional  Bogoliubov
transformation of the new  the operators $a_k$ and $a_k^\dagger$
to diagonalize the Hamiltonian \cite{bogol}.
\begin{eqnarray}
a_k &=& \frac{1}{\sqrt{1 - A_k^2}} (b_k + A_k b_{-k}^\dagger),
\nonumber \\
a_k^\dagger &=& \frac{1}{\sqrt{1 - A_k^2}} (b_k^\dagger + A_k
b_{-k}).
\label{21}
\end{eqnarray}
It is clear that $b_k$ and $b_k^\dagger$ satisfy the same commutator
relations as $a_k$ and $a_k^\dagger$.
 Then,  the Hamiltonian in Eq. (\ref{7})
becomes
\begin{eqnarray}
\hat{H} &=&  \frac{1}{2} n^2 V g
 +   \sum_{k\neq
0}\frac{1}{1-A_k^2}\left[(\vep_k + n g)A_k^2
 + n g A_k \right]
 \nonumber \\
&  +& \frac{1}{2}\sum_{k\neq 0}\frac{1}{1-A_k^2} \left[(\vep_k
+ n g)(1+ A_k^2) + 2 n g A_k  \right]
  (b_k^\dagger b_k +
b_{-k}^\dagger b_{-k})
 \nonumber \\
&  +& \frac{1}{2}\sum_{k\neq 0}\frac{1}{1-A_k^2}
\left[2(\vep_k + n g)A_k +  n g(1+ A_k^2)  \right]
(b_k^\dagger b_{-k}^\dagger + b_{k} b_{-k})
 \nonumber \\
& +& \frac{n g}{\sqrt{N}}\sum_{k \neq 0}
\frac{(1+A_k)}{\sqrt{1-A_k^2}}
 (b_k + b_{-k}^\dagger) \gamma_k.
\label{23}
\end{eqnarray}
$A_k$ is chosen to make the off-diagonal term vanish.
\begin{equation}
2(\vep_k + n g)A_k +  n g(1+ A_k^2) = 0, \label{25}
\end{equation}
and  we obtain $A_k(=A_{-k})$ as
\begin{equation}
A_k = \frac{ E_k - (\vep_k + ng)}{n g}, \label{27}
\end{equation}
where
\begin{equation}
E_k = \sqrt{(\vep_k + n g)^2 - (n g)^2}. \label{29}
\end{equation}
The $E_{k}$ is the Bogoliubov form
of the dispersion relation.

Substituting  $A_k$ into the Hamiltonian in Eq. (\ref{23}), we
obtain the following compact form of the Hamiltonian
\begin{eqnarray}
\hat{H}   &=&  \frac{1}{2} n^2 V g
+  \sum_{k\neq0}
\frac{1}{1-A_k^2}\left[(\vep_k + n g)A_k^2
 + n g A_k \right]
  \nonumber \\
& +& \sum_{k\neq 0} E_k b_k^\dagger b_k
  + \sum_{k\neq 0} G_k (b_k +
b_{-k}^\dagger), \label{31}
\end{eqnarray}
where
\begin{equation}
G_k = \frac{ng}{\sqrt{N}}\frac{(1+A_k)\gamma_k}{\sqrt{1-A_k^2}}.
\label{33}
\end{equation}
The first three terms in the right hand side of Eq. (\ref{31})
are already known and the last one is the new
contribution beyond the  textbook results \cite{huan,fett,abri}.
We can regard the system as a collection of quantum mechanical
 harmonic oscillators exposed to additional forces given
by linear terms.
The linear terms $b_k$ and $b_k^\dagger$ can be
made to eliminate by another linear transformation of the form
\begin{eqnarray}
b_k &=& c_k + \alpha_k,
\nonumber \\
b_k^\dagger &=& c_k^\dagger + \alpha_{-k}, \label{35}
\end{eqnarray}
where the new operators $c_k$ and $c_k^\dagger$ satisfy the bosonic
commutator relations. Note that $\alpha_k$ commutes with $c_k$ and
$\alpha_k^\dagger=\alpha_{-k}$ since $b_k$ and $b_k^\dagger$ are
bosonic operators.

Substituting the new operators in Eq. (\ref{35}) into Eq.
(\ref{31}), we can rewrite the Hamiltonian in Eq. (\ref{31}) as
\begin{eqnarray}
\hat{H} &=&  \frac{1}{2} n^2 V g + \sum_{k\neq
0}\frac{1}{1-A_k^2}\left[(\vep_k + n g)A_k^2
 + n g A_k \right]
\nonumber \\
& +& \sum_{k\neq 0} \left( E_k c_k^\dagger c_k + E_k
\alpha_{-k} \alpha_k + 2 G_k \alpha_k \right)
 \nonumber \\
& +& \sum_{k\neq 0} \left[ (E_k \alpha_{-k} + G_k) c_k +
(E_k \alpha_{k} + G_{-k}) c_{k}^\dagger \right].
\label{37}
\end{eqnarray}
We choose $\alpha_k $  to make the linear terms vanish as
\begin{equation}
\alpha_k = -\frac{G_{-k}}{E_k}.
\label{375}
\end{equation}
Then, the Hamiltonian in Eq. (\ref{37}) is now
 diagonalized in terms of the new operators as follows
\begin{eqnarray}
\hat{H} &=&  \frac{1}{2} n^2 V g
+ \sum_{k\neq 0}\frac{1}{1-A_k^2}\left[(\vep_k + n g)A_k^2
 + n g A_k \right]
 - \sum_{k \neq 0} \frac{G_k G_{-k}}{E_k}
% \nonumber \\
%& +&
+ \sum_{k\neq 0} E_k
c_k^\dagger c_k
\nonumber \\
&=& \hat{H}_g + \sum_{k\neq 0} E_k n_k.
\label{39}
\end{eqnarray}
${\hat H}_g$ creates the well-known ground state energy density \cite{huan,fett,abri}
 but we will not discuss it here.
\begin{equation}
\frac{E_g}{V} = \frac{2\pi\hbar^2 a n^2}{m}
\left[ 1+ \frac{128}{15}\left( \frac{n a^3}{\pi} \right)^{1/2} + \dots \right].
\label{40}
\end{equation}

%%%%%%%%%%%%%%%%%%%%%%%%%%%%%%%%%%
\section{condensate traction}
%%%%%%%%%%%%%%%%%%%%%%%%%%%%%%%%%%

The condensate fraction is defined as $(N-\sum_{k \neq 0} n_k)/N$.
We obtain the particle distribution $n_k$ of the dilute system
from Eqs. (\ref{21})  and (\ref{35}) as
\begin{eqnarray}
n_k &=& \langle 0| a_k^\dagger a_k |0 \rangle
 \nonumber \\
&=& \frac{A_k^2}{1-A_k^2}
+ \frac{(1+A_k)^2}{1-A_k^2} \langle 0|\alpha_{-k} \alpha_k |0 \rangle
\nonumber \\
&=& \frac{1}{2}
\left( \frac{\vep_k + ng}{E_k} - 1 \right)
 +\frac{1}{N} \frac{n^2 g^2\vep_k^2}{E_k^4}
\langle 0| \gamma_k
\gamma_{-k} |0 \rangle.
\label{43}
\end{eqnarray}
Note that $(1+A_k)/(1-A_k)=\vep_k/E_k. $
The magnitude of $n g\vep_k/E_k^2$ is order of 1.
The first term is shown in the textbooks and produces the well-known $c_1$
in Eq. (\ref{1}).
On the other hand, the relation between the second term and $n_k$ will create $c_2$.

The dominant part of the unknown quantity
 $\langle 0| \gamma_k \gamma_{-k}  |0 \rangle$
is obtained in the following way.
Applying the definition of $\gamma_k$ in Eq. (\ref{5}),
the ground state average is written as
\begin{equation}
\langle 0| \gamma_k \gamma_{-k} |0 \rangle
= \langle 0|
\sum_{q,q'\neq 0} a_{k+q}^\dagger a_{q}
a_{-k+q'}^\dagger a_{q'} |0 \rangle.
\label{44}
\end{equation}
The summation is composed of the two terms as
$\sum_{q,q'\neq 0}
= \sum_{q'=k+q}   +  \sum_{q'\neq k+q}. $
     It is known that the dominant contribution arises
when a particle interacts with itself and it belongs to the term
 ${\bf q'} = {\bf k} + {\bf q}$.
Then, we can separate the dominant
contribution into two parts ${\bf q}={\bf q'}$ and ${\bf q} \neq
{\bf q'}$ again, but
the term for ${\bf q} \neq {\bf q'}$ vanishes using the argument of
random phase approximation. Therefore, we have
\begin{eqnarray}
\langle 0| \gamma_k \gamma_{-k} |0 \rangle
& \simeq& \langle 0|\sum_{q'=k+q}  a_{k+q}^\dagger
 a_{q} a_{-k+q'}^\dagger a_{q'} |0 \rangle
\nonumber \\
&=& \langle 0| \sum_{q=q'\neq 0} a_{q'}^\dagger a_{q} a_{q}^\dagger
a_{q'} |0 \rangle
  + (k{\rm -dependent~minor~terms}).
\label{45}
\end{eqnarray}
%From now on we do not carry the $(k{\rm -dependent~minor~terms})$
%in $\langle 0| \gamma_k \gamma_{-k} |0 \rangle$.

Within this approximation,
using the $n_k$ in Eq. (\ref{43}), we can write Eq. (\ref{45}) as
\begin{eqnarray}
\langle 0| \gamma_k \gamma_{-k} |0 \rangle
&\simeq& \sum_{q\neq 0} n_q^2
\nonumber \\
&=& \sum_{q\neq 0}
 \left[ \frac{1}{2} \left( \frac{\vep_q + n g}
{E_q} - 1 \right) \right]^2
 + (k{\rm -dependent~minor~terms})
 \nonumber \\
&=& N\left[ \left( \pi - \frac{ 8 }{3} \right) \eta + \mathcal{O}(\eta^2) \right],
\label{46}
\end{eqnarray}
where $\eta=\sqrt{n a^3/\pi} \ll 1$. We put $\hbar=m=1$,
and $\sum_q = (N/2\pi^2 n)\int q^2 dq$ for convenience.
The cutoff of the k-dependent minor terms
is at most order of $\eta^2$ or smaller.
The effective range of the cutoff is judged the  by an  iteration method
self-consistently.

 Substituting $\langle 0| \gamma_k \gamma_{-k}
|0 \rangle$ in Eq. (\ref{46}) into Eq. (\ref{43}), we obtain a new
higher order $n_k$ as
\begin{equation}
n_k = \frac{1}{2} \left( \frac{\vep_k + n g}{E_k} - 1 \right)
+ \left( \pi - \frac{8}{3} \right)\frac{n^2 g^2 {\vep_k}^2 }{E_k^4}
 \eta + \mathcal{O}(\eta^2).
\label{47}
\end{equation}
Now, let us back the new $n_k$ into Eq. (\ref{46}) to obtain
a new higher order $\langle 0| \gamma_k \gamma_{-k} |0 \rangle$, too.
\begin{eqnarray}
\langle 0| \gamma_k \gamma_{-k} |0 \rangle
&=& \sum_{q\neq 0}
 \left[ \frac{1}{2} \left( \frac{\vep_q + n g}
{E_q} - 1 \right) + \left( \pi - \frac{8}{3} \right)
\frac{n^2 g^2 {\vep_q}^2 }{E_q^4}
 \eta +\mathcal{O}(\eta^2)\right]^2
 \nonumber \\
 &=&N \left[\left( \pi - \frac{8}{3}\right)\eta
+ 2\left(\pi - \frac{8}{3}\right)\left( \frac{10}{3}-\pi\right)\eta^2
+ \frac{\pi}{16} \left(\pi - \frac{8}{3}\right)^2 \eta^3
+ \mathcal{O}(\eta^3)\right].
 \label{49}
\end{eqnarray}
Therefore, we can trust up to $\eta^2$ term in
$\langle 0| \gamma_k \gamma_{-k} |0 \rangle$
with the cutoff.

Finally, substituting Eq. (\ref{49}) into Eq. (\ref{43}),
we obtain the next higher order two coefficients $c_2$ and $c_3$.
The particle depletion from the zero momentum condensate,
$(N-N_0)/N$,  is
\begin{eqnarray}
 \frac{1}{N}\sum_{k \neq 0} n_k
 &=&\frac{1}{2N}\sum_{k \neq 0}\left( \frac{\vep_k + n g}{E_k}
- 1\right)
%\nonumber \\ & &
+  \frac{1}{N}\sum_{k \neq 0}\frac{{n^2 g^2\vep_k}^2}{E_k^4}
\left[\left( \pi - \frac{8}{3}\right) \eta + 2\left(\pi -
\frac{8}{3}\right) \left( \frac{10}{3}-\pi \right)\eta^2 + \mathcal{O}(\eta^3)
\right]
\nonumber\\
&=& \frac{8}{3} \eta + 2 \pi \left( \pi -\frac{8}{3}\right)
\eta^2 + 4\pi \left( \pi -\frac{8}{3}\right)\left( \frac{10}{3}-\pi \right)\eta^3
+ \mathcal{O}(\eta^4)
\nonumber\\
&=& \frac{8}{3\sqrt{\pi}}(n a^3)^{1/2}
+ 2 \left( \pi -\frac{8}{3}\right) n a^3
+ \frac{4}{\sqrt{\pi}} \left( \pi -\frac{8}{3}\right)\left( \frac{10}{3}-\pi \right)(n a^3)^{3/2}+\mathcal{O}((na^3)^2).
 \label{61}
\end{eqnarray}
Therefore, finally, we obtain $c_2=2(\pi - 8/3)$ and
$c_3=(4/\sqrt{\pi}) \left( \pi -8/3\right)\left( 10/3-\pi \right)$.
%This approximation is effective for the major term only.

%%%%%%%%%%%%%%%%%%%%%%%%%%%%%%%%%%
\section{summary}
%%%%%%%%%%%%%%%%%%%%%%%%%%%%%%%%%%

We expanded the Hamiltonian of a homogeneous and interacting Bose
system up to the $\sqrt{N_0}$ terms, and applied the canonical
transformation twice to find the average value of the higher order
terms. An iteration method was applied to make up the cutoff.
Two higher order coefficients $c_2$ and $c_3$ were obtained analytically.

%%%%%%%%%%%%%%%%%%%%%%%%%%%%%%%%%%%%%%%
%\acknowledgments
%%%%%%%%%%%%%%%%%%%%%%%%%%%%%%%%%%%%%%%
%The author dedicates this paper to Dr. H.S. Noh
%who passed away in 2006.

%%%%%%%%%%%%%%%%%%%%%%%%%%%%%%%%%%%%%%
%%%%%   REFERENCES   %%%%%%%%%
%%%%%%%%%%%%%%%%%%%%%%%%%%%%%%%%%%%%%%%
{}
%%%%%%%%%%%%%%%%%%%%%%%%%%%%%%%%%%%%%%%
\end{document}